# High Performance Architecture for Flow-Table Lookup in SDN on FPGA

Rashid Hatami[a], Hossein Bahramgiri[a] and Ahmad Khonsari[b]

[a]Maleke Ashtar University of Technology, Tehran, Iran
[b]Tehran University and Institute for Studies in Theoretical Physics and Mathematics (IPM), Tehran, Iran

**Abstract**

We propose Range-based Ternary Search Tree (RTST), a tree-based approach for flow-table lookup in SDN network. RTST builds upon flow-tables in SDN switches to provide a fast lookup among flows. We present a parallel multi-pipeline architecture for implementing RTST that benefits from high throughput and low latency. The proposed RTST and architecture achieve a memory efficiency of 1 byte of memory for each byte of flow. We also present a set of techniques to support dynamic updates. Experimental results show that RTST can be used to improve the performance of flow-lookup. It achieves a throughput of 670 Million Packets Per Second (MPPS), for a 1 K 15-tuple flow-table, on a state-of-the-art FPGA.

## 1. Introduction

Consider an SDN network where the data plane forwards the traffic according to the decisions made by the control plane. The control plane is managed by a logically centralized controller to instruct data plane. The forwarding is determined by rules that data plane apply to packets. The OpenFlow [1, 2] can be used as a viable interface for SDN to manage the network traffic between the control and data plane. An OpenFlow switch uses one or more flow tables to perform its functionality. One of the main challenges is the flow table lookup operation during which the data plane must identify the corresponding rule for each incoming packet quickly, according to the desired match. It compares a search string (prefix of the incoming packet fields) with a table of stored rules (flow-table), and returns the matching rule. In order to improve performance in SDN networks, data plane must provide a fast lookup on the flow-table and so benefit from memory efficiency and dynamically updatable architecture. Since switch memory may be rather needed to deal with the always growing number of services or to guarantee network performance [3], it is important to consider memory efficiency and updatable architecture. Some proposals focus on Ternary Content Addressable Memory (TCAM) [4-6], Bit Vector (BV)-based schemes [7, 8], or the tree-based approaches [9, 10]. TCAM is not scalable with respect to the flow-table size; also it is a scarce, expensive, and power-hungry resource [11]. Additionally, the time of adding new rules with the random priority is significantly higher than with the single priority (approximately 6 times) [12]. The rules in TCAM are stored from top to bottom in decreasing priority, and for this reason inserting a new TCAM entry often involves rearranging existing entries. This yields high overhead for each TCAM update. The BV algorithms can provide a high throughput at the cost of low memory efficiency [10]. In [9], two algorithms are presented, one is Binary Search Tree (BST) with focusing on memory efficiency and the other is 2-3 tree with focusing on incremental update. The main drawback of BST data structure is the difficulty in supporting updates because the BST needs to be rebuilt for new updates. The 2-3 tree may incur significant cost in terms of resources because there is need to store child pointer at each node. Internet applications require some hardware to perform frequent updates and adapt to different processings [13, 14]. Because it is prohibitively expensive to reconstruct an optimal architecture repeatedly for timely updates, many solutions have been proposed on FPGA [7]. In this paper, we introduce a high throughput and memory efficient flow-table lookup with RTST. To perform a fast lookup, RTST reduces the search space of flow-table. RTST has no need to be rebuild for new updates and therefore, this obtains efficient update. Moreover, RTST takes advantage of memory efficiency by means of the pointer elimination technique. We present a parallel multi-pipeline architecture on FPGA. This architecture is high throughput, low latency and also supports efficient dynamic update. The main contributions of this paper are as follows.

- We propose RTST as a data structure to provide fast lookup on the flow tables.
- We present an RTST-based pipeline architecture which benefits from high throughput even if the length of the packet header is scaled up.
- We employ an optimization technique to achieve the high memory efficiency of 1 byte of memory per byte of flow.
- We present a set of techniques for supporting dynamic updates, including modifying, deleting and inserting operations on the flow-tables.

The rest of the paper is organized as follows: Section 2 introduces the background and problem definition. Section 3 details RTST design. Section 4 describes the proposed architecture. Section 5 presents update techniques on this architecture. Section 6 describes our design scalability and the performance evaluations. Section 7 concludes the paper.

## 2. Background and Problem Definition

### 2.1. Background

The forwarding decisions are routine necessity that are taken locally at each SDN switch. First, the SDN controller generates the rules according to the request of the operator. Second, these rules install on the SDN switches to be applied to packets traversing on the network. In the OpenFlow, each rule translated in the flow-table as a flow. The flow indicate that network devices (typically switches) how to process or forward packets. Each flow is associated with a specific action. According to the decision policy, each action includes the updating of the flow, forward to destination and queue assignment [2].

### 2.2. Problem Definition

The problem of flow-table lookup is defined as follows: given a flow-table $F$ consists of $N$ flow, design a high-throughput lookup engine which also supports memory efficiency and dynamic updates. It compares the incoming packet fields against a flow-table, and returns the matching rule. A flow entry is matched with a packet if only all the header fields of the packet are matched with the match fields of the flow. Among all the matched flows, the highest priority flow is considered. The lookup operation on the flow-table requires a large number of field for matching. For example, in the recent specification of the OpenFlow protocol [2], the number of fields are 15 in the packet header, which consists of 356 bits. Two kinds of matches are allowed in each field in a flow: prefix match and exact match. In a prefix match, the header field should be a prefix of one. In an exact match, the header field of the packet should exactly match the flow field. In the flow-table lookup, only the source address (SA) and destination address (DA) fields require the prefix match, while all the other fields require the exact match. To achieve a high performance lookup engine, there are three existing challenges that need to be resolved:

- *Large scale:* lookup upon the large number of flows is one of the main bottlenecks in large-scale networks because it need to high processing.
- *Memory efficiency:* The increase in the size of flow-table necessitates the high memory-efficient lookup methods.
- *Dynamic update:* Flows may be dynamically changed and updated and this affects the performance. Dynamic update carried out in the flow-table includes three operations: 1) modification, 2) deletion and 3) insertion.

## 3. RTST Design

In this section, we propose RTST, a special tree-based data structure that is focusing in high memory efficiency and quick lookup. In our design, order of the implemented tree is three because the hardware implementation results show that the order greater than three are very difficult to be efficiently implemented. The order greater than three inflicts the poor uniformity in the data structure and so construct a wide verity of fat nodes. In this case, the wide memory access is required for each fat node [9] and yielding significantly memory overhead caused by the unused nodes and increasing the number of the bits for storing pointers as a reference at each node. RTST includes the following properties:
- The tree is a type of range-based ternary tree.
- The tree includes two different types of node: 3-node and leaf-node. All of the nodes contain up to two data fields, left and right data field, where the value of the right data field is greater than the left one. Figure 1 shows types of nodes in RTST.
- The 3-node contains up to two data fields and three pointers. Each 3-node references to three children, include left, middle and right, by means of pointers. One child contains values less than the left data field, the other one contains values between the two data fields, and the last one contains values greater than the right data field.
- All of the leaf-nodes contain one or two data fields and are at the lowest level.

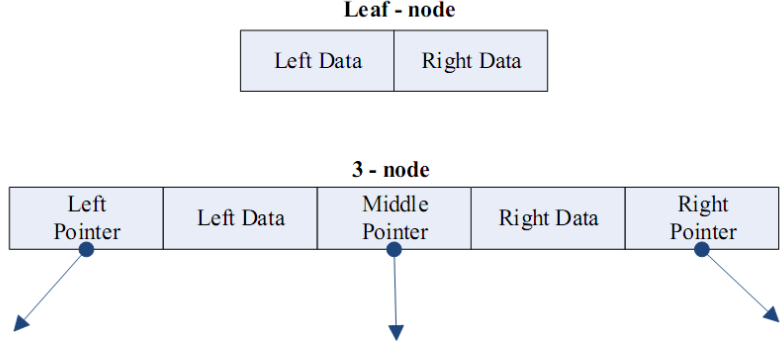

Figure 1: Different types of nodes in RTST.

A complete RTST is a tree in which all the levels are fully occupied except possibly the last level. In a complete RTST, each element can be found in $1 + \log_3^N$ operations, at the worst case, where $N$ is the total number of nodes in the tree.

In the first step, the root of tree, which consists of two values, must be determined. The values location of root can be calculated as follows,

$$L_L = \left\lfloor \frac{1}{3} N \right\rfloor \quad (1)$$

$$L_R = \left\lfloor \frac{2}{3} N \right\rfloor, \quad (2)$$

where $N$ is the total number of values, $L_L$ is the location of the left data, and $L_R$ is the location of the right data. After unveiling the root, three sub-trees, $T_L$, $T_M$ and $T_R$, are obtained. The $T_L$ is the left sub-tree which has locations $0 \leq T_L < L_L$, the $T_M$ is the middle sub-tree which has locations $L_L < T_M < L_R$, and the $T_R$ is the right sub-tree that has locations $L_R < T_R \leq N$. For each sub-tree, this procedure is repeated recursively. Let $n$ be the number of levels of the RTST and $\gamma$ be the number of nodes in the last level. The total number of nodes in each level, excluding the last level, is $3^{n-1}$ and $\gamma = N - \sum_{i=1}^{n} 3^{i-1}$. The steps to build an RTST are described in Algorithm 1.

---

**Algorithm 1:** Complete RTST
**Data:** List $A$ consist of $N$ sorted elements
**Result:** Complete RTST
1. $L_L = \left\lfloor \frac{1}{3} N \right\rfloor, L_L = \left\lfloor \frac{2}{3} N \right\rfloor$
2. Root = [$L_L$, $L_R$]
3. Pick elements ($A[L_L]$, $A[L_R]$)
4. Left sub-tree = complete RTST (left of $L_L$ sub-list);
5. Middle sub-tree = complete RTST (middle of $L_L$ and $L_R$ sub-list);
6. Right sub-tree = complete RTST (right of $L_R$ sub-list);

---

We divide a flow-table into $k$ groups to achieve significant performance that will be described in subsection 6.3 Two groups of flows are said to be disjoint if and only if any two flows in the groups do not overlap with each other. An RTST is built for each group of disjoint flows. The

RTST need not to be rebuild for inserting new flows and the flows can be added in the tree in any order. To add a new flow *F* into RTST, compare *F* with node's data fields and move down the correct path while the leaf node is not reached. The following cases are defined to insert a new flow upon RTST:
- In the leaf node, if the leaf node has one data item, then insert *F* into the leaf node as the left or right data field.
- If the leaf node has two data items and the parent node has one data item, then compare the two data items of the leaf node with *F*. Afterwards, select the middle value from the comparison and insert it in the parent node. If leaf node was the left child, then move the data in the parent node to the right as the right data field and insert the middle value in the parent node as the left data field. If leaf node was the right child, then the middle value is inserted in the parent node as the right data field and the two left-over value become two leaf nodes.
- If the leaf and parent nodes are full, then develop the middle value of the leaf node to its parent node. Afterward, develop the middle value of the current node while the parent node is full and current node is not root. If current node is root, then create a new root node from the middle value to its parent. Finally, update the left and right data fields of the new root.
- 

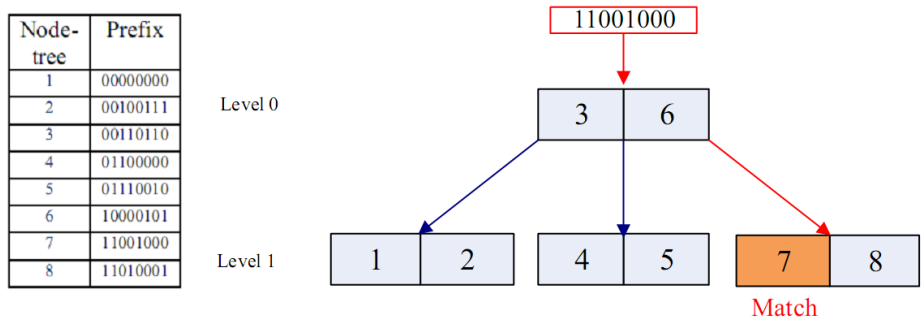

Figure 2: A sample group of disjoint prefixes and its corresponding complete RTST.

Figure 2 shows a simple structure of RTST in which a disjoint prefixes of a group is implemented. Depending on the comparison result at each node, the traversing path is determined and search is performed by traversing left, middle or right. If the incoming prefix is smaller than the left data field, then the traversal leads to the left child through the left pointer, and if the prefix is greater than the right data field, then the traversal leads to the right child through the right pointer and the traversal leads to the middle child, otherwise. The input value is compared with the root, and the decision about the traversing path is made based on the data values of each node. For example, assume that a packet with a prefix $P_p = 11001000$ arrives. $P_p$ is compared with node values *00110110* and *10000101*, yielding no match and a "right" result. Thus, the packet traverses to the right sub-tree. The comparison with the input value in this node results in a "match", which is the matched result of the tree.

*3.1 Improving the memory efficiency by pointer elimination*

The RTST is fast to lookup the stored data and flexible to update new data. Each 3-node in RTST contains two data and three pointers. To build the RTST, each child pointer is need to be stored as a reference at each node. These pointers may incur cost in terms of memory consumption. Thus, the memory efficiency can be improved if these pointer are eliminated.

In order to eliminate pointers, must provide a logical relationship between parent and child nodes. The nodes in each level are accommodated in a contiguous memory block. Let $i$ denotes the index of node $x$; then $3i$ is the index of the left child of node $x$, and its middle child and right child are located at $3i + 1$ and $3i + 2$ indexes, respectively. The address of the next node could simply be calculated based on the current node address. Thus, child pointers are calculated explicitly on-the-fly and there is no need to store child pointers at each node. The memory allocation and address calculation are described in Figure 3. Consider a node with address $A_{i,j}$, where $i$ and $j$ are level and the node number of tree, respectively; to access the left child, $(3 \times A_{i,j})$ should be calculated, and $((3 \times A_{i,j}) + 1)$ and $((3 \times A_{i,j}) + 2)$ should be calculated in order to access the middle and right children, respectively.

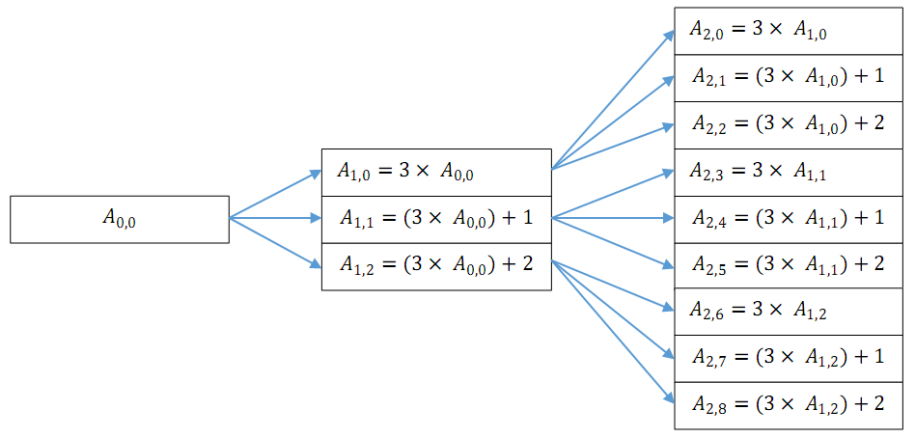

Figure 3: Pointer elimination in RTST.

## 4. Architecture

### 4.1. Overall Architecture

Pipelining is an important technique that is used to increase the throughput. In the case of the ideal pipeline (without hazard), it takes one clock cycle for each operation. Throughput and delay are two criteria used to examine the performance of a pipeline. They are related but not identical. There are two approaches to increase the performance of pipeline architecture. The first one is to reduce the delay in each stage. The second one is to increase the number of parallel operations in each stage. We use both approaches to improve the throughput of our design. The overall architecture is shown in Figure 4. In our multi-pipeline architecture, there are $k$ pipeline, one for each group of the flow table. To implement RTST on the pipeline, each level of the tree is mapped upon a separate pipeline stage. The number of pipeline stages is determined by the height of the tree, and the height depends on the number of elements of each group. The proposed architecture is composed of two sections: Source and Destination Search Tree (SST and DST). DST includes all the flow fields except the source field. For the incoming packet $P$, the prefix of the source address is searched on SST. If the prefix of the source address matches upon SST, packet P will traverse on DST. The prefix of the destination address and other fields are searched at the same time.

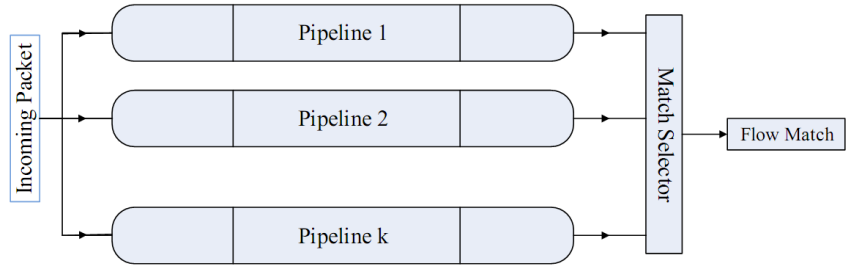

Figure 4: The overall architecture.

## 4.2. RTST architecture

The block diagrams of the pipeline architecture and a single stage are shown in Figure 5. The pipeline architecture is dual-line and each stage from that includes logical circuits and a memory block. Logical circuits are applied to perform the comparison operation and send the result to the next stage. Memory block is dual port so that two data can be read/write per every clock cycle.
In each stage of the pipeline, there are three data forwarded from the previous stage: 1) input prefix (which is compared with the content of each entry in the memory to determine the matching status), 2) memory address, and 3) ready signal. The forwarded memory address is used to retrieve the content of each entry in the memory that is the left and right data of the node. Note that the access time of the memory is a major reason for delay on each stage. It can affects the total performance. In our design, each entry in the memory includes two data elements. In this case, two data elements are available with one memory access, hence the access time delay will be decreased. When a matching is made, the ready-signal is set and memory access in the stages is turned off.

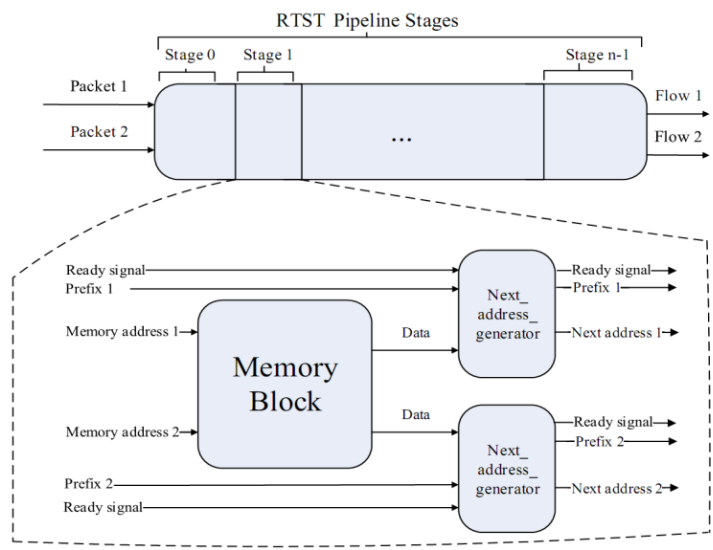

Figure 5: the proposed architecture for implementing RTST.

Figure 6 shows the block diagram of the next address generator. The next address and the memory access are calculated in parallel so that there is no extra delay for generating the next address. In the output of the comparator, four states are possible: 1) input prefix is smaller than the left data, for which the priority encoder provides the $3 * A_i$. 2) The input prefix is between the left and right data, for which the priority encoder provides the $3 * A_{i+1}$. 3) The input prefix is greater than the right data, for which the priority encoder provides the $3 * A_{i+2}$. 4) The match is made either with the left or right data. In this case, the ready-signal is established to refer to the next stage. When a match is made, searching in the subsequent stages is unnecessary. Hence, memory access in those subsequent stages can be turned off and the previous result can simply be forwarded.

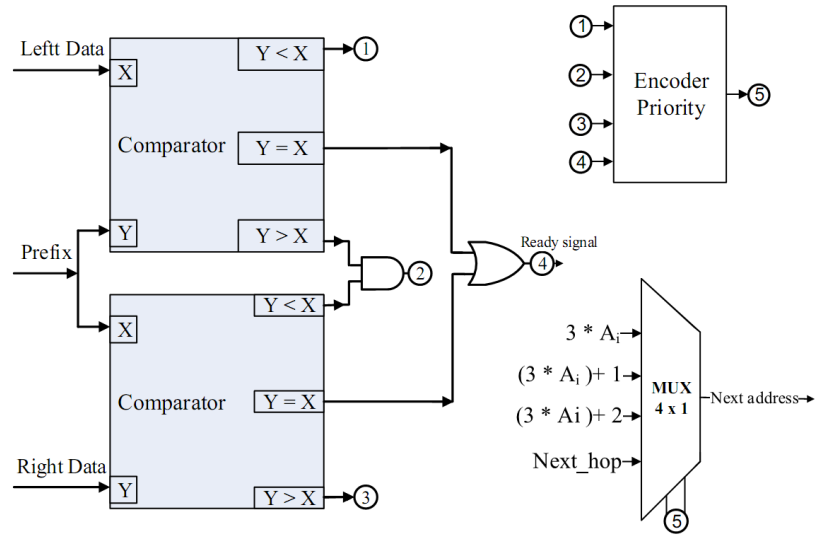

Figure 6: The block diagram of next address generator.

## 5. Dynamic updates

Dynamic update for the flow-table is a problem that requires the hardware to adapt to frequent updates during run-time. Given a flow-table consisting of $N$ flow entries $\{F_i |\ i = 0,\ 1,...,\ N-1\}$, the updates include three operations: 1) modification, 2) deletion, and 3) insertion.

- **modification:** The primary purpose of the modification is to change flows information in the flow-tables. For the modification requests, if a matching entry exists in the flow-table, the fields of this entry are updated with the new value from the request. Given a flow $F$, search $F$ in the flow-table, locate $i \in \{0, 1, ..., N-1\}$ where $F_i = F$; change flow $F_i$ in the flow-table into flow $F$.

- **deletion:** The primary purpose of deletion is to remove the flows in the flow-tables. For the deletion requests, if a matching entry exists in the table, it must be deleted. Given a flow $F$, search $F$ in the flow-table, locate $i \in \{0, 1, ..., N-1\}$ where $F_i = F$; remove $F_i$ completely from the flow-table.

- **insertion:** The primary purpose of insert is to add the flows in the flow-tables. Two flow entries are overlap if a single packet may match both, and both entries have the same priority. If an overlap conflict exists between an existing flow entry and the insert request, the insertion must be refused. In the non-overlapping case, insert the flow entry in the flow-table. Given a flow $F$, search $F$ in the flow-table, if $\forall i \in \{0, 1, ..., N-1\}, F_i \neq F$; then insert $F_i$ into the flow-table.

The first step of the update operations is always $F$ check, which reports whether the $F$ of the new flow exists in the current flow-table. After $F$ check is completed, we present our main ideas as follows.

## 5.1 Modification

After $F$ check, assume $F$ exists in the flow-table; hence $\exists i \in \{0, 1, ..., N-1\}$ such that $F_i = F$. Flow modification can be performed by inserting a write bubble, as introduced in [15]. This is shown in Figure 7. There is one write bubble in each stage that stores the update information. Each write bubble consists of two entries. Each entry includes: 1) the memory address to be updated in the next stage, 2) the new content, and 3) a write enable bit. When a flow modification is initiated, the memory content of the write bubble in each stage is modified and a write bubble is inserted into the pipeline. The write bubble will modify the memory location in the next stage when the write enable bit is set. Since the memory is dual-ported, two nodes can be simultaneously modified at each stage.

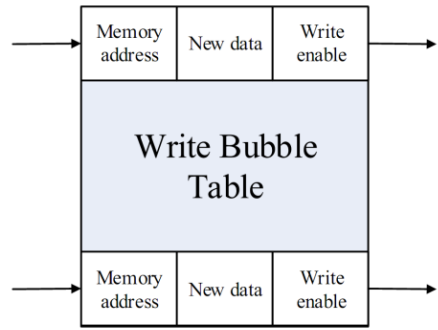

Figure 7: dual-ported write bubble table.

## 5.2. Deletion

After $F$ check, assume $F$ exists in the table; hence $\exists i \in \{0, 1, ..., N-1\}$ such that $F_i = F$. For deletion, we keep a "valid" bit for each node. A valid bit is a binary digit indicating the validity of a node. A node is valid only if its corresponding valid bit is set to '1', and is invalid if its corresponding valid bit is set to '0'. To delete a flow, we reset its corresponding valid bits to '0'. An invalid flow is not comparable for any match result. We show an example for deletion in Figure 8. In this example, initially $F8$ is invalid and the other one is valid. If we want to delete the $F4$ and $F7$, the valid bits corresponding to $F4$ and $F7$ are reset to '0'.

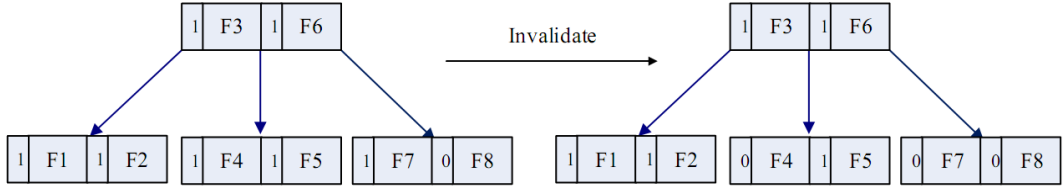

Figure 8: An example for deletion.

*5.3. Insertion*

After *F* check, assume that *F* does not exist in the table; hence $\forall i \in \{0, 1, ..., N-1\}, Fi \neq F$. For insertion, we insert a new flow by modifying an invalid node or adding a new node. In other words, we reuse the location of the invalid flows to insert the new flows. If there isn't invalid node, create a new node upon RTST. Algorithm used to insert upon RTST is the bottom-up approach. The first step is to find the parent *p* of the newly node *n* that will be inserted on RTST. There are two cases: 1) the *p* node has only one or two children (the parent node has one or two invalid children) and 2) the *p* node has three children (the parent node has not got the invalid child). In the first case, the valid bit of a child node is set to '1' and *n* is inserted as the appropriate child of *p*. In the second case, if *n* is inserted as the appropriate child of *p*, *p* has four children, violating the property of a RTST. Hence, an internal node *m* is created according to the mechanism of adding a new flow upon RTST. In the worst case, the insertion causes changes in at most two nodes at each level of the RTST.

## 6. Performance Evaluation

*6.1. Scalability*

The great challenge to develop scalable solutions is to support higher throughput, larger flow-tables and more packet header fields. Our design benefits from high clock rate even if the length of each packet header is scaled up, as well as fast lookup, when the number of flows is increased. The use of the complete RTST data structure provides a linear storage complexity in our design. Since each level of RTST is mapped to a pipeline stage, the number of stages is determined by the height of RTST. The structure of the RTST never has a height greater than $\lfloor log_3^N \rfloor$ and provides a fast lookup on flow-tables. Another advantage of this approach is that the lookup latency does not depend on the field length. In the proposed architecture, implementation is simple and the amount of logic resources is small. We can expect considerable resource savings in hardware implementation.

*6.2. Experimental Setup*

We conducted experiments to evaluate the performance of our design, including the FPGA prototype of the architecture. The architecture has been implemented in VHDL, using Xilinx ISE

Design Suite 14.2, targeting the Virtex 6 XC6VLX760 FFG1760-2 FPGA [16]. The chip contains 118, 560 logic slices, 1200 I/O pins, 26Mb BRAM (720 RAMB36 blocks). A Configurable Logic Block (CLB) on this FPGA consists of 2 slices, each having 4 LUTs and 8 flip-flop. Throughput and latency are reported using post-place-and-route results. We also generated random flows for OpenFlow-tables (d = 15, L = 356), where d is the total number of fields for the flow and *L* is the total number of bits for the input packet header, in order to prototype our design. The number of flows in a flow-table is chosen from 128 to 1K.

### *6.3. Scalability of Latency*

Latency is the processing delay of a single packet when no dynamic update is performed. The latency performance with respect to various values of N is shown in Figure 9. We used the prefix partitioning [5] to divide a given flow-table into *k* groups of disjoint flows. Increasing the number of groups can be ill-suited in hardware implementation. For this reason, we can achieve a permissible number of groups by $h = log_3^{(\frac{N}{k})}$, where *h* is the height of the search tree. As can be seen, the prefix partitioning approach benefits from better latency performance in comparison with the no prefix partitioning approach.

### *6.4. Comparison of the Throughput and Latency with the State-of-the-art*

We compare the throughput and latency performance of our approach with the existing hardware accelerators in Figure 10. All the implementations and our approach support 1K flows and each flow includes 15-fieldes. The TCAM can match packets in a single clock cycle [17]. Hence, the throughput and latency performance of TCAM are estimated based on the assumption that packets can be matched within a single clock cycle and clock rate can be at 360MHz. In the FSBV [18], it is assumed that 7 pipeline stages are employed and the clock rate can be sustained at 167MHz. For StrideBV [19], it is assumed that the single-pipeline implementation consisting of 89 stages running at 105 MHz is employed. For the decision-tree-based implementation [10], it is assumed that the 16-stage pipeline is employed and clock rate can be at 125MHz. In the BV-based pipelined architecture [8], the 2-dimensional pipelined architecture is employed, which, for a flow-table consisting of N flows, and an L-bit packet header, requires N rows and L columns while the clock rate can be sustained at 324MHz. As can be seen, the TCAM has the lowest latency and StrideBV introduces the highest latency. Compared to the StrideBV, decision-tree and BV-based pipelined architecture, our design achieves better latency. Our architecture achieves high throughput compared to the other approaches. Our architecture also achieves high clock rate (337MHz) and throughput (674MPPS).

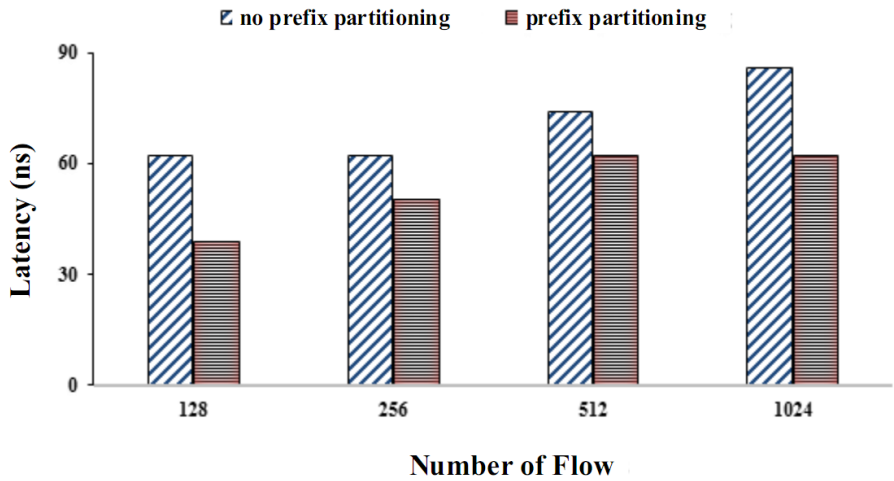

Figure 9: Latency for each N = 128, 256, 512, and 1024.

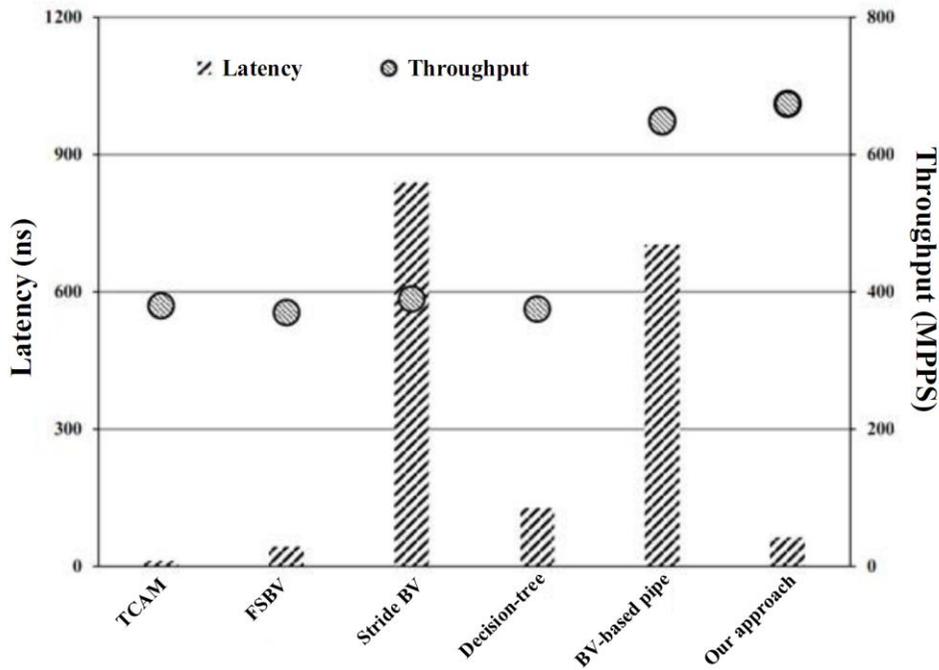

Figure 10: Comparing throughput and latency

## 6.5. Memory Efficiency

The memory efficiency is defined as the amount of memory required to store one flow entry. We compared our design with TCAM [17], FSBV [18] and StrideBV [19] with respect to the average memory requirement per flow in table 1. The BV-based pipelined architecture [8] is unclear for the amount of memory requirement, but usually the BV algorithms can provide high throughput at the cost of low memory efficiency. As can be seen, our design consumed 44.5 bytes per flow, achieving a memory efficiency of 1 byte of memory for each byte of flow. TCAM also consumes 44.5 bytes per flow, but is not scalable in terms of circuit area, as compared to SRAMs [20].

Table 1: Comparison of the memory efficiency.

| Approach | Memory req. | |
| --- | --- | --- |
| | L = 104 bits | L = 356 bits |
| TCAM [17] | 13 Bytes/Flow | 44.5 Bytes/Flow |
| FSBV [18] | 29 Bytes/Flow | 99.23 Bytes/Flow |
| Stride BV [19] | 156 Bytes/Flow | 534 Bytes/Flow |
| Our approach | 13 Bytes/Flow | 44.5 Bytes/Flow |

# 7. Conclusion

In this paper, we proposed RTST as a data structure to achieve fast lookup for the flow-tables. We also presented a parallel multi-pipeline architecture to sustain both high throughput and low latency. This architecture benefits from high clock rate and fast lookup, as well as supporting dynamic updates. The proposed architecture is simple and the amount of logic resource is small. We could also achieve a memory efficiency of 1 byte of memory for each byte of the flow entry. Experimental results showed that for a 1 K 15-tuple flow-table, a state-of-the-art FPGA could sustain a throughput of 670 MPPS. With these advantages, our algorithm and design can be used to improve the performance of flow-lookup.